\documentstyle[12pt,epsfig]{article}
\begin{document}
\title{Quantum nondemolition measurements in a Paul trap}
\author{ A. Camacho
\thanks{email: acamacho@aip.de} \\
Astrophysikalisches Institut Potsdam. \\
An der Sternwarte 16, D--14482 Potsdam, Germany.}

\date{}
\maketitle

\begin{abstract} 
In this work a family of quantum nondemolition variables for the case of a particle caught in a Paul trap
is obtained. Afterwards, in the context of the so called restricted path integral formalism,
a continuous measuring process for this family of parameters is considered, and then the
corresponding propagators are calculated. In other words, the time evolution
of a particle in a Paul trap, when the corresponding quantum nondemolition
parameter is being continuously monitored, is deduced.
The probabilities associated with the possible measurement outputs are also obtained,
and in this way new theoretical results emerge, which could allow us to confront
the predictions of this restricted path integral formalism with the readouts of some future experiments.
\end{abstract}

\section{Introduction}

One of the more long--standing conundrums in modern physics is the so called
quantum measurement problem (QMP) [1], the one from the very outset of quantum theo\-ry (QT) has provoked
a deep interest by its unusual and paradoxical characte\-ris\-tics. Nowadays the attempts in the quest for a solution of
this conceptual difficulty have been spawned by theoretical and also by practical reasons.
On practical grounds, for instance, the necessity of detecting very small
displacements in the case of gravitational--wave antenna [2], or the attempts
to obtain high sensitivity in parametric transducers [3] (a topic closely related to
the design of gravitational radiation detectors), requires the analysis of QMP.

On the theoretical side, QMP plays a fundamental role in the understanding of the
foundations of QT [1], and therefore a better comprehension of this issue
is closely related to the advancement of QT.

Concerning QMP it is noteworthy
to co\-mment that one of the most interesting topics in this field is the so called quantum nondemolition
measuring process [4], in which certain class of observables may be measured repeatedly with
arbitrary precision. Of course, the dynamical evolution of the corresponding system
limits this class of observables. Clearly this restriction is a direct consequence of the unavoidable back reaction
of the measuring device upon the measured system.
The fundamental idea behind a quantum nondemolition measurement (QNDM) is to monitor a variable
such that the unavoidable disturbance of the conjugate one does not perturb
the time evolution of the chosen parameter [3].

On experimental grounds QNDM is very promising [5], for instance, in the dy\-na\-mics of the interaction between measuring device
and a mechanical oscillator the most important hurdles, that currently
impede the achievement of the corresponding quantum regime, have already been identified, and the
exploration of the quantum behavior in the context of macroscopic mechanical oscillators
could bring in the near future decisive results [6].

In order to confront the theoretical predictions of RPIF against experimental
outputs the necessity of performing continuous measurements upon a quantum system is imperative. In this respect we must mention
that the current technology allows us to carry out repetitive measurements on a single quantum system. For instance,
it is already possible [7] to confine and observe an individual electron in a Penning trap, or we can now also have a trapped single atom and 
observe its \-in\-ter\-ac\-tion\- with a radiation field, for example, by means of a laser fluorescence [8]. 
In the case of the so called Paul trap, which has led to the construction of a mass spectrometer [9], 
a ion is trapped employing a high-frequency electric quadrupole field [10]. As is known, this idea 
can also be extended to the case of neutral atoms, laser cooled and stopped atoms are confined in a magnetic quadrupole trap 
formed by two opposed, separated, coaxial current loops [11]. 

The aforementioned experiments open
the possibility of confronting, in a near future, the\- theo\-re\-ti\-cal\- predictions of some formalisms that claim to describe the interaction between measuring apparatus 
and measured system against experimental results. 
One of these formalisms is the so called Restricted Path--Integral Formalism 
(RPIF) [12]. The main idea in this approach is the restriction by means of a weight functional of the integration domain of the 
path--integral that renders the \-corres\-pond\-ing\- propagator of the analyzed system, when one or more of its parameters 
are subject to a continuous measurement process. 

Let us explain this point a little bit better, and suppose that we have a particle which has a one--dimensional movement. 
The amplitude $\hat{U}(q'', q')$ for this particle to move from the point $q'$ to the point $q''$ is called propagator,
and it is given by Feynman [13]

\begin{equation}
\hat{U}(q'', q') = \int d[q]exp\left({i\over \hbar}S[q]\right),
\end{equation}

\noindent here we must integrate over all the possible trajectories $q(t)$, and $S[q]$ is the action of the system, which is 
defined as

\begin{equation}
S[q] = \int_{t'}^{t''}dtL(q, \dot{q}).
\end{equation}

Let us now suppose that we perform a continuous measurement of the position of this particle, 
such that we obtain as result of this measurement process a certain output $a(t)$. In other words, the measurement process gives the value $a(t)$ 
for the \-coor\-di\-na\-te $q(t)$ at each time $t$, and this output has associated a certain error $\Delta a$, which is determined by the 
experimental resolution of the measuring device. The amplitude $\hat{U}_{[a]}(q'', q')$ can be now thought of as a probability amplitude for the continuous measurement process to give the result $a(t)$. 
Taking the square modulus of this amplitude allows us to find the probability density for different measurement outputs.

Clearly, the integration in the Feynman path--integral should be restricted to those trajectories that match with the experimental output. 
RPIF states that this condition can be introduced by means of a weight functional $\tilde{\omega}_{[a]}$ [14]. 
This means that expression (1) 
becomes now under a continuous measurement process

\begin{equation}
\hat{U}_{[a]} = \int d[q]\tilde{\omega}_{[a]}exp(iS[q]).
\end{equation}

The more probable the trajectory $[q]$ is, according to the output $a$, the bigger that $\tilde{\omega}_{[a]}$ becomes [12], 
this means that the value of $\tilde{\omega}_{[a]}$ is approximately one for all trajectories, $[q]$, that agree with the measurement 
output $a$ and it is almost 0 for those that do not match with the result of the experiment. 

Clearly, the weight functional contains 
all the information about the interaction between measuring device and measured system, and the problems that in this context appear are two: 

(i) The concrete form of the weight functional $\tilde{\omega}_{[a]}$ depends on the measuring device [15], in other words, 
the involved experimental constructions determine these weight functionals. We must determine $\tilde{\omega}_{[a]}$ starting from the 
knowledge that we have of the measuring device, a non--trivial problem. 

(ii) If we wish an analytical expression for $\hat{U}_{[a]}$, then the resulting weight functional 
must render an analytically handleable functional integral.

In this work we will consider the case of a particle caught in a Paul trap, and the
nonlinear differential equation, the one defines the corresponding quantum nondemolition
variables, will be solved. As solution a family of quantum nondemolition parameters will be obtained. Afterwards, a continuous measuring process for one of these parameters
will be considered, and the associated propagators will be calculated. Finally,
the probabilities of the different measurement outputs are obtained.
\bigskip
\bigskip

\section{Quantum nondemolition variables in a Paul trap}
\bigskip
\bigskip

It is already known that in the case of a particle in a Paul trap we have a harmonic oscillator, the one possesses a frequency equal to $\bar{U} - \bar{V}cos(\omega t)$, 
being $\bar{U}$, $\bar{V}$, and $\omega$ constants which depend on the electric quadrupole field used to trap the 
particle. The motion equations of an electrically charged particle caught in a Paul trap are [10]
(here we will assume that ions are injected in the $y$--direction and that there is electric field only along the $x$-- and $z$--coordinates)

\begin{equation}
\ddot{x}(t) + {e\over mr^2}[\bar{U} - \bar{V}cos(\omega t)]x(t) = 0,
\end{equation}

\begin{equation}
\ddot{z}(t) - {e\over mr^2}[\bar{U} - \bar{V}cos(\omega t)]z(t) = 0.
\end{equation}

\noindent here $e$ is the electric charge of the particle, $m$ its mass, and $2r$ the distance between the electrodes that conform part of the 
experimental apparatus. The solutions to these equations are the so called Mathieu functions [16]. 

Let us now first consider the motion only along the $x$--axis. This is no restriction at all, because the complete motion can be separated in 
two independent motions. Thus, our starting point is the following Lagrangian 

\begin{equation}
L = {1\over 2}m\dot{x}^2(t) - {1\over 2}m[U - Vcos(\omega t)]x^2(t).
\end{equation}
\noindent here we have introduced the following definitions: $U = {e\over mr^2}\bar{U}$ and $V = {e\over mr^2}\bar{V}$. 

In the case of a harmonic oscillator with a time dependent frequency a quantum nondemolition (QND) variable
may be found in the form [12]

\begin{equation}
A = \rho p + \sigma q,
\end{equation}

\noindent where $p$ and $q$ denote the momentum and position, respectively, and $\rho$ and
$\sigma$ satisfy the following differential equation [12]

\begin{equation}
{df\over dt} = {f^2\over m} + m[U - Vcos(\omega t)],
\end{equation}

\noindent here $f = {\rho\over\sigma}$.

Let us now consider the following function:

\begin{equation}
F = -{m\over x(t)}{dx\over dt},
\end{equation}

\noindent where $x(t)$ is any of the possible solutions to

\begin{equation}
\ddot{x}(t) + [U - Vcos(\omega t)]x(t) = 0.
\end{equation}

It is straightforward to check that $F$ is a solution to (8). Hence, with the solutions of the
motion equation we may easily find QND--variables. In other words, we have found a
family of QND--parameters for the case of a particle in a Paul trap.

\begin{equation}
A(t) = \sigma(t)\left[-m{1\over x(t)}{dx\over dt}q + p\right].
\end{equation}

If we set $V = 0$, then we recover, from expression (9), the situation already known of a harmonic oscillator with time independent frequency [12].
Indeed, in that case, $x(t) = -\cos(\sqrt{U}t)$, and then $\rho/\sigma = m\sqrt{U}\tan{(\sqrt{U}t)}$,
where $\sqrt{U} = \omega$ is the corresponding frequency of oscillation.

\bigskip
\bigskip

\section{Continuous measurements, propagators, and pro\-ba\-bility densities}
\bigskip
\bigskip

Let us now consider the situation in which our QND--variable, $A$, is being conti\-nuous\-ly
monitored, here we set $\sigma(t) = 1$. According to RPIF [12] the corresponding
propagator (when the measurement output reads $a(t)$) is given by

{\setlength\arraycolsep{2pt}\begin{eqnarray}
\hat{U}_{[a]}= \int_{q'}^{q''}d[q]d[p]\exp\left\{{i\over \hbar}
\int_{t'}^{t''}\left({p^2\over 2m} + {1\over 2m}[U - Vcos(\omega t)]q^2\right)dt\right\}\tilde{\omega}_{[a]}.
\end{eqnarray}} 

As was mentioned at the end of the first section, the whole interaction between measuring
apparatus and measured system is contained in $\tilde{\omega}_{[a]}$. At this point we must choose
a particular weight functional, and it will be a gaussian one

\begin{equation} 
\tilde{\omega}_{[a]}= \exp\left\{-{1\over T\Delta a^2}\int_{t'}^{t''}[A(t) - a(t)]^2dt\right\}.
\end{equation} 

The reasons for this choice lie on the fact that the results coming from a
Heaveside weight functional [14] and those
coming from a gaussian one [17] coincide up to the order of magnitude.
These last remarks allow us to consider a gaussian weight functional as an approximation of the correct expression.
Hence, it will be supposed that the weight functional of our measuring device has
precisely this gaussian form. We may wonder if this is not an unphysical assumption,
and in favor of this argument we may comment that recently it has been proved that
there are measuring apparatuses which show this kind of behavior [18].

Therefore, expression (12) becomes

{\setlength\arraycolsep{2pt}\begin{eqnarray}
\hat{U}_{[a]}= \int_{q'}^{q''}d[q]d[p]\exp\left\{{i\over \hbar}
\int_{t'}^{t''}\left({p^2\over 2m} + {U - Vcos(\omega t)\over 2m}q^2 + {i\hbar\over T\Delta a^2}[A - a]^2\right)dt\right\}.
\end{eqnarray}}

This last path integral is gaussian in $p$ and $q$, and therefore, it can be easily calculated [19]

{\setlength\arraycolsep{2pt}\begin{eqnarray}
\hat{U}_{[a]}= \exp\left\{-{1\over T\Delta a^2}\int_{t'}^{t''}a^2dt\right\}\nonumber\\
\times\exp\left\{{T\Delta a^2 -2im\hbar\over 2m^2\hbar\gamma}
\int_{t'}^{t''}{(a{\dot{x}\over x})^2\left[2m^2\hbar\alpha+ imT\Delta a^2\beta\right]\over\alpha^2 + {T^2\Delta a^4\over 4m^2\hbar^2}\beta^2}\right\},
\end{eqnarray}}

\noindent where we have introduced three new definitions, namely,
$\alpha = ({\dot{x}\over x})^2 + U - Vcos(\omega t)$,
$\beta = U - Vcos(\omega t)$, and $\gamma = 4m^2\hbar^2 +T^2\Delta a^4$.

The probability densities, associated with the different measurement outputs, read
(according to the expression $P_{[a]} = \vert \hat{U}_{[a]}\vert^2$)

{\setlength\arraycolsep{2pt}\begin{eqnarray}
P_{[a]}= \exp\left\{-{2\over T\Delta a^2}\int_{t'}^{t''}a^2dt\right\}
\exp\left\{{T\Delta a^2\over \gamma}
\int_{t'}^{t''}{(a{\dot{x}\over x})^2\left[({\dot{x}\over x})^2 + 2\beta\right]\over\alpha^2 + {T^2\Delta a^4\over 4m^2\hbar^2}\beta^2}\right\}.
\end{eqnarray}}
\bigskip
\bigskip

\section{Conclusions}
\bigskip
\bigskip

In this work we have considered the case of a particle caught in a Paul trap,
and, after solving the corresponding nonlinear differential equation, a family of quantum nondemolition variables has been obtained, expression (11). Afterwards,
a continuous quantum measurement for an element of this family was considered,
and, along the ideas of the so called restricted path integral, the co\-rres\-ponding
propagator has been calculated, expression (15). Finally, the associated probability densities
were derived, expression (16). The present work complements a previous paper in which a quantum demolition
measurement for a particle in a Paul trap was analyzed [20].

Clearly, we may notice in our last equation that there is no standard quantum limit, in other words, we may measure $A$ with
an arbitrarily small error, and in consequence all the necessary information can be extracted.

Looking at (16) we may notice that in the limit $\Delta a \rightarrow 0$
we obtain $P_{[a]}\rightarrow 0$. In other words, in the limit of
very precise measurements all the possible readouts have the same probability density. 
This last fact is a quantum feature, indeed, in the nonquantum case only the solution
to the classical motion equations has a nonvanishing probability. This last remark does not mean 
that a strong disturbance of the corresponding observable is present. 
Indeed, we may find this kind of behavior even in a much simpler situation, namely, in the case of a harmonic 
oscillator the limit of very precise measurements ($\Delta a \rightarrow 0$) renders 
also the result $P_{[a]}\rightarrow 0$ (see equation (6.32) in [12]). This cha\-rac\-teristic is
also present in the situation when a quantum nondemolition variable, for the case of a particle
moving in an inhomogeneous gra\-vi\-tational field, is monitored in the limit of very small
intrumental error [21]. The opposite case, namely, the limit of rough measurements, $\Delta a\rightarrow \infty$, renders 
also the result $P_{[a]}\rightarrow 0$.

Of course, we could have a very small experimental error, but the case $\Delta a \rightarrow 0$ 
is an idealization, and this limit has to be understood in the sense that if the 
ex\-pe\-ri\-mental resolution is much smaller than all the relevant physical variables, then 
we could expect to have a probability independent of the measurement outputs. Ex\-pe\-ri\-mentally 
this case could be a very difficult one, consider, for example, the current experimental
resolution in the case of Paul or Penning traps [10], which lies very far from this idealization.

There are already some theoretical results [22, 23] that provide a framework
which could allow us to confront the predictions of RPIF against some possible future
ex\-pe\-riments.
Nevertheless, the present work renders new theoretical predictions coming from RPIF,
which in the future could be confronted with experimental readouts. For example,
expression (16) predicts a very particular dependence (on the involved measurement readouts) for the ratio of the probability densities
associated with two different experimental results. Indeed, if $b(t) \not = a(t)$, then its is readily seen that $P_{[a]}/P_{[b]} = g(a^2 - b^2)$,
where $g$ is a real function

{\setlength\arraycolsep{2pt}\begin{eqnarray}
P_{[a]}/P_{[b]}= \exp\left\{-{2\over T\Delta a^2}\int_{t'}^{t''}\left(a^2 - b^2\right)dt\right\}\nonumber\\
\times\exp\left\{{T\Delta a^2\over \gamma}
\int_{t'}^{t''}{(a^2 - b^2)({\dot{x}\over x})^2\left[({\dot{x}\over x})^2 + 2\beta\right]\over\alpha^2 + {T^2\Delta a^4\over 4m^2\hbar^2}\beta^2}\right\}.
\end{eqnarray}}

To finish, let us comment an additional characteristic of expression (16). We know that there
are some QNDMs in which an absolute limit may appear. For instance,
if the linear momentum of a free particle is monitored, this aforementioned limit may emerge,
when the instantaneous measurement of position of the test particle, before and after the monitoring
of the linear momentum, is done (see page 99 of reference [12]).
Clearly, position and linear momentum
are canonical conjugate variables to each other, and that is why this absolute limit appears. At this
point we may wonder why (16) shows no absolute limit, in our case, position, before and after monitoring of $A(t)$, has also been instantaneously measured, i.e.,
$q'$ and $q''$ are present from the very begining in our mathematical expressions, see (12).

The answer to this question stems from the fact that in our case the measured
quantity, $A = \sigma(t)[-m{1\over x(t)}{dx\over dt}q + p]$, is not the canonical conjugate variable of position, $q$.
\bigskip
\bigskip
\bigskip

\Large{\bf Acknowledgments}\normalsize
\bigskip

The author would like to thank A. A. Cuevas--Sosa for his 
help, and D.-E. Liebscher for the fruitful discussions on the subject. 
The hospitality of the Astrophysikali\-sches Institut Potsdam is also kindly acknowledged. 
This work was supported by CONACYT (M\'exico) Posdoctoral Grant No. 983023.

\end{document}